\documentclass[useAMS,usenatbib]{mnras}

\usepackage{amsmath,amssymb,amsfonts}
\usepackage{bm}
\usepackage{float}
\usepackage{graphicx}
\usepackage{color}
\usepackage[varg]{txfonts}
\usepackage{enumerate}
\usepackage{enumitem}
\usepackage{indentfirst}

\usepackage{ulem}

\newcommand{\Deltaimax}{$\Delta_{i,\rm max}$}
\newcommand{\imax}{$i_{\rm max}$}
\newcommand{\gap}{{\it gap} }

\begin{document}
	
  \title[gap in satellite mass]{The Formation of M101-alike Galaxies in the Cold Dark Matter Model}
  
  \author[D. Zhang et al.] 
  {Dali Zhang$^{1,2}$, Yu Luo$^{1,2,4}$\thanks{E-mail: luoyu@pmo.ac.cn}, Xi Kang$^{3,1}$\thanks{E-mail: kangxi@zju.edu.cn}  and Han Qu$^{1,2}$\\
  	$^1$Purple Mountain Observatory, No. 10 Yuanhua Road, Nanjing 210033, China\\
    $^2$School of Astronomy and Space Sciences, University of Science and Technology of China, Hefei 230026, China\\
    $^3$Zhejiang University-Purple Mountain Observatory Joint Research Center for Astronomy, Zhejiang University, Hangzhou 310027, China\\
    $^4$National Basic Science Data Center, Zhongguancun South 4th Street, Beijing 100190, China.\\
  }

  %\date{Manuscript Version: Oct 2019}
  
  %\pagerange{\pageref{firstpage}--\pageref{lastpage}} \pubyear{2021}
  
  \maketitle
  
  \label{firstpage}
  
  \begin{abstract}
  The population of satellite galaxies in a host galaxy is a combination of the cumulative accretion of subhaloes and their associated star formation efficiencies, therefore, the luminosity distribution of satellites provides valuable information of both dark matter properties and star formation physics. Recently, the luminosity function of satellites in nearby Milky Way-mass galaxies has been well measured to satellites as faint as Leo I with $M_{V} \sim -8$. In addition to the finding of the diversity in the satellite luminosity functions, it has been noticed that there is a big gap among the magnitude of satellites in some host galaxies, such as M101, where the gap is around $5$ in magnitude, noticeably larger than the prediction from the halo abundance matching method. The reason of this gap is still unknown. In this paper, we use a semi-analytical model of galaxy formation, combined with high-resolution N-body simulation, to investigate the probability and origin of such big gap in M101-alike galaxies.  We found that, although M101 analogues are very rare with probability of $\sim 0.1\%-0.2\% $ in the local universe, their formation is a natural outcome of the CDM model. The gap in magnitude is mainly due to the mass of the accreted subhaloes, not from the stochastic star formation in them. We also found that the gap is correlated with the total satellite mass and host halo mass. By tracing the formation history of M101 type galaxies, we find that they likely formed after $z \sim 1$ due to the newly accreted bright satellites. The gap is not in a stable state, and it will disappear in ~7 Gyr due to mergers of bright satellites with the central galaxy. % or "due to mergers ... with central galaxies"
    
  \end{abstract}


  \section{Introduction}
  In the $\Lambda$ cold dark matter model, the formation of dark matter structure is hierarchical, in the pattern that small haloes form first, and they subsequently merge to form larger ones \citep{Fre12}. 
  The relics of the merging haloes and their stellar components are called as satellite galaxies. %I am sorry, but I do not understand this last sentence
  Massive galaxy usually sits at the center of the dark matter halo where the gas cooling and mergers with satellites are in progress. The evolution of satellites is more or less passive, as the cold gas is continuously consumed by star formation, and additional gas supply is cut off by ram-pressure stripping  \citep[e.g.][]{Gun72, Mar03, May06, Kang08, Nic11, Luo16, Eme16, Sim17}. From this picture, it is natural to expect that galaxy properties such as luminosity, star formation and color, are systematically different between central and satellite galaxies. 
  
  One prominent feature of a normal galaxy group/cluster is that there is a luminosity gap between the central galaxy and the most luminous satellite. It was firstly noted that some galaxy groups display a large gap in magnitude with $\Delta M>2$ \citep{Pon94}, and they were named as "fossil group". Later studies \citep[e.g.,][]{Jon03} suggested that the gap is an indicator of halo age such that fossil groups form early and the most massive satellites have merged with the central galaxy, thus producing a large gap in their stellar mass or magnitude. Further studies \citep{Dea13,Zhang19} also found that for hosts with a given stellar mass for the central galaxy, those with small gap form at a later epoch and have larger halo mass than those with a big gap. Thus, hosts with a small gap between the central and the most massive satellite tend to have more satellites, and this effect should be taken into account when comparing the distribution of satellites around central galaxies with similar stellar mass.

  Up to now, most studies have focused only on the luminosity gap between the central and the most massive satellite. Little attention is paid to the luminosity gap between satellites themselves %What? I don't understand... do you maybe mean "left alone its...?
  \citep[although see][on the velocity gap between satellites in the Milky Way]{Jiang15, Kang16}. In recent years, the search of satellite population around nearby % local and nearby have basically the same meaning, so, choose one
  Milky Way-mass galaxies has obtained a complete % be careful in using the word "complete". Observers might intend it in a different way....
  sample of faint dwarfs with magnitude down to -8 in V band \citep[e.g.][]{Bro14, Wang15, Wet16, Gra17,Ben19,  Ben20,Crn19, Chi13, Sme17, Mar16, McC18}. These data reveal that there is a large scatter between the luminosity functions of satellites in different host galaxies, %Again, I don't understand the sentence. Do you mean that there are many luminosity functions that differ from each other?
  and it is also found that 
  for some galaxies, such as M101, there is a big gap ($\sim -5$) between the magnitude of the satellite population, noticeably larger than the gap in other MW-mass galaxies. 
  %This gap is larger than the halo-to-halo scatter predicted from the abundance matching model (Geha et al. 2007, Bennet et al. 2019). 
  
  At a first sight, it seems not surprising to find a big gap between satellites. It is well known that, although the mass function of subhaloes is nearly universal with a weak dependence on the host mass \citep[e.g.][]{Spr01, Gao04}, there is a large scatter among halos with given mass \citep[e.g.][and reference therein]{Jiang14}, so the gap is possibly rooted in the mass distribution of accreted subhaloes.  However, it was shown \citep{Geh17, Ben19} that the gap in the observational data, such as M101, is larger than the predictions from the halo abundance matching model in the cold dark matter model.  Meanwhile, \citep{Geh17, Mao20} also mentioned that the assumption in the abundance matching is too simplistic, thus caution should be taken when interpreting the comparison of the gap between observational data and simulation. For these reasons, it is still not clear to which extent the magnitude gap between satellites challenges the cold dark matter model.
  
%  For example, they use using the  they claim that the large gap in M101 challenge the galaxy formation model. Can the gap between satellites tell us something on galaxy formation physical or halo formation history? Why the gap in M101 is so large and what makes the difference between M101 and MW, or other galaxy system in near universe?
  
  %In the $\Lambda$CDM model, the structure and substructure are expected to be self-similar. The mass profile of all halos look alike when scaled properly, regardless of mass \citep{Nav96, Nav97}, and the mass function of subhalos in each halo has similar distribution when it is scaled to the mass of the host \citep{Moo99, Wang12}. So the different gaps between satellites can also reflect the different structure and galaxy formation process on small scales.

  In this paper, we employ the semi-analytical model of of \citet{Luo16} % either you name the SAM, or otherwise use the article "a", not "the". Your choice.
  and the Millennium-II simulation(MS-II, \citep{Boy09}) to study the gap in more detail. Compared with the halo abundance matching model, the semi-analytical model follows more physical process which can in principle produce more diverse star formation efficiencies in low-mass satellites, and the simulation used here has a higher resolution than that used by \citet{Geh17} to resolve faint satellites. We focus on the gap in M101-alike galaxies %btw, you may consider to use "like" rather than "alike"
  to constrain the probability of finding M101-alike galaxies from the simulation, % probability of what? Stated like this, it doesn't make sense.
  and for those with big gap, we investigate which physical process, the scatter in the subhalo mass at accretion or stochastic star formation efficiency in the satellite, is the main factor in producing the gap. %again....it's confused. You investigate what process between the scatter in the subhalo mass at accretion and the star formation efficiency is the main responsible for the gap? If so, just say it. Anyway, none of them is a physical process. Star formation is a physical process, not its efficiency. And a scatter definitely not.

  \section{Model and Sample Selection}
  In this work, we use the semi-analytical galaxy formation model of \citet{Luo16} %Finally!!!! You should name it earlier!
  run on the MS-II to produce model galaxies. The MS-II is a dark matter only cosmological simulation with 2160$^3$ particles with box size of 100Mpc/h. By rescaling the cosmological parameters from the WMAP1 to the WMAP7 cosmology (a method developed by \citet{Ang10}), the box size of MS-II has changed to 104.311 Mpc/h, and the particle mass has changed from 6.9$\times 10^6M_\odot/h$ to 8.5$\times 10^6M_\odot/h$. The resolution of this simulation is high enough to resolve faint satellites down to $M_V$=-5 \citep{Guo15}. The semi-analytical model (SAM) of \citet{Luo16} is a resolution independent %what? what do you mean with "resolution independent"? That's absolutely impossible!
  version based on the Munich galaxy formation model: L-Galaxies  \citep[e.g.][]{Kau93, Kau99, Spr01, Cro06, De07, Guo11, Guo13, Fu10, Fu13, Hen15}. For more detail about the model, we refer the readers to the references above.
  
  In the L-Galaxies model, galaxies are classified into three types: Type 0, Type1 and Type2 . Type 0 galaxies are those located at the centre of Friends-of-Friends (FOF) halo groups and are called centrals (host galaxies) of these groups. Both Type 1 and Type 2 galaxies are satellite galaxies in the model. A Type 1 galaxy is located at the center of a subhalo, while type 2 represent the so-called "orphan galaxies", i.e. satellites without resolved subhaloes. Type 2 is usually the descendant of a Type 1 galaxy. The halos/subhaloes containing at least 20 bound particles are catalogued in the MS-II simulation\citep{Spr05, Boy09}.

\begin{figure}
  	\includegraphics[width=\linewidth]{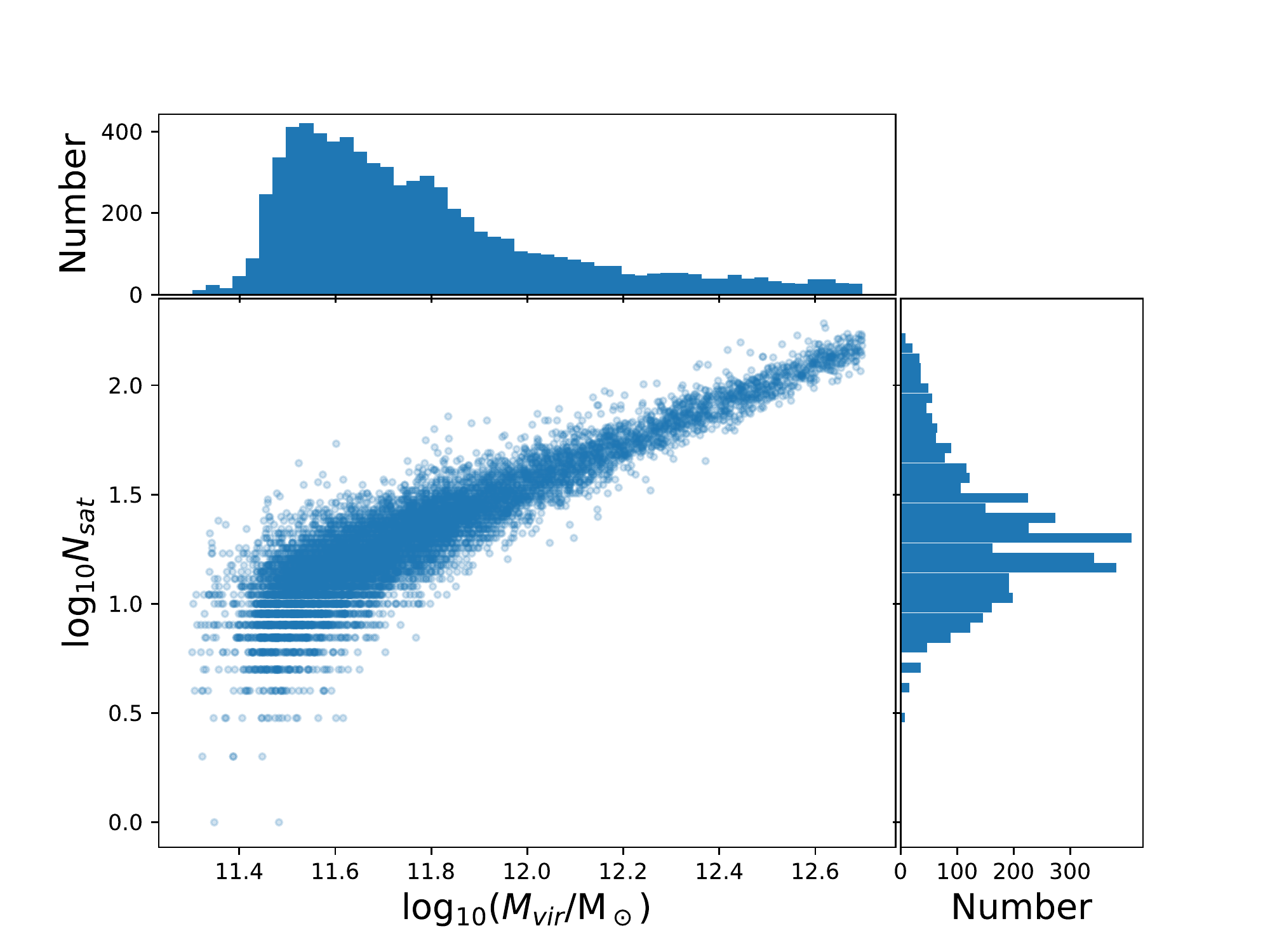}
  	\caption{The relation between halo virial mass and number of satellites in the model. The main panel represents the mass-satellite number relation, and the upper and right panel represent the number distribution of mass and satellites number. }
  	\label{fig:gap-num}
  \end{figure}

  To investigate the large-gap galaxy like M101, we select central galaxies in our SAM %as far as I remember you never state the acronym SAM above. Do it
  catalog with $M_V$ between -20 and -22,  according to the typical V band magnitudes of MW, M101, and other MW-mass galaxies. We obtain 9756 galaxies with a V-band magnitude cut of -8 for satellites within a 2D projection distance of 250 kpc from the central galaxy. In Fig~\ref{fig:gap-num}, we show the virial mass %did the referee ask what virial mass you refer to? If not, it's ok.
  and the number of satellites in our sample. The figure shows that the number of satellites has a good linear correlation with halo virial mass. Most of central galaxies are located in halos with virial mass about $10^{11.5} \sim 10^{12} M_\odot$, and these galaxies are expected to have dozens of satellites within their virial radii.
  
\begin{figure}
  	\includegraphics[width=\linewidth]{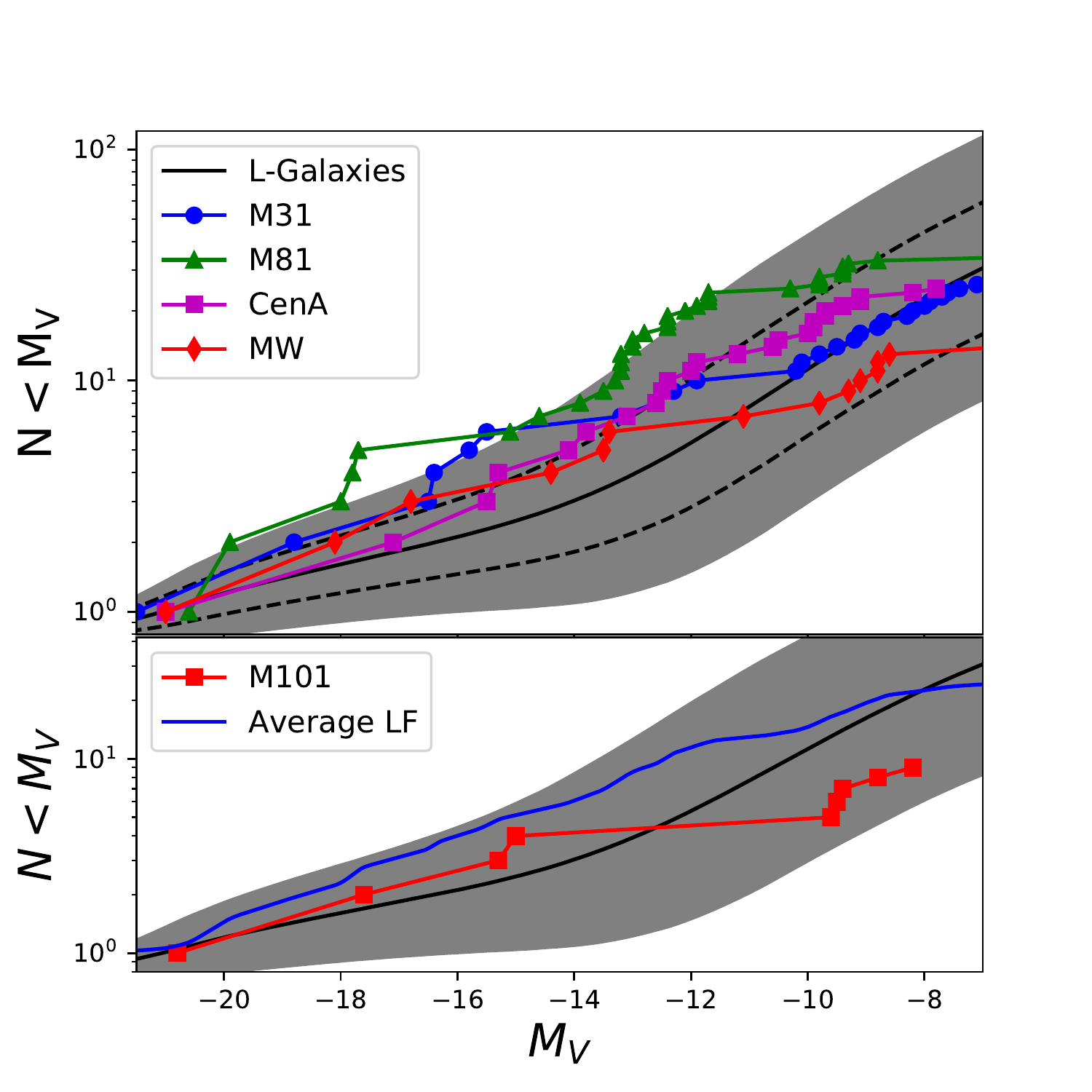}
  	\caption{The cumulative luminosity function (LF) of satellites in several (including center galaxies) Milky Way-mass galaxies, out to a projected radius of 250 kpc. The black line is the average value of model galaxies, with black dashed lines that represent the 1$\sigma$ region, and the gray shadow that represents the 2$\sigma$ region.  The color lines with different markers in the upper panel are several observational galaxy systems. The average LF %another acronym not specified earlier
	of the 4 galaxies is shown as a blue line in the lower panel.  The data are from \citet{Crn19} for Cen A (represented by magenta squares), \citet{Chi13} and \citet{Sme17} for M81 r(epresented by green squares), \citet{Mar16} \& \citet{McC18} for M31 r(epresented by blue squares), and \citet{McC12} for the MW (represented by the yellow squares). }
  	\label{fig:sam-range}
  \end{figure}

  In Fig~\ref{fig:sam-range}, we show the cumulative luminosity function of satellites of our SAM sample, and compare it with several Milky Way analogous galaxies in the local universe. Basically, the satellite luminosity distribution of our SAM sample (gray shadow) is similar to the observational data. Furthermore, in the lower panel, we compare the average satellite luminosity distribution of MW analogous (blue line) and M101 (red dots) with the SAM sample. The average satellites number distribution of the SAM sample (black line) shows a double power law relation with the magnitude. There are more fainter satellites and less brighter ones in our SAM sample than in MW analogous, consistent with the results found by \citet{Geh17}.  In particular, M101 has a similar distribution compared with other galaxies or SAM samples at the bright end, but with less satellites at the faint end. Obviously, M101 shows a quite large gap between the third ($M_{V, 3}=-15$) and the fourth ($M_{V,4}=-9.6$) satellites, with a magnitude gap of 5.4. 
  
 \section{Results}
   \subsection{Are M101-alike galaxies rare in the local universe?}
   Since only M101 has a big gap in its satellite luminosity among these Milky Way analogous, in this section, we investigate whether the big gap in M101 is rare in our universe, by searching M101-alike galaxies in our SAM data which can reproduce the average satellite distribution of Milky Way analogous.
  
   To define the gap in a galaxy system, we use $i$ to represent the $i$th galaxy in the order of magnitude: e.g. $i=0$ represents the central galaxy and $i=1$ represents the brightest satellite. The magnitude gap of galaxies are represented as $\Delta$, and $\Delta_i$ represents the gap between the $i$th and $(i+1)$th luminous galaxies. In this work, we focus on the largest gap in magnitude between the satellites (hereafter, we use \gap to represent the largest magnitude gap), whose value is represented as \Deltaimax. The order of \gap is represented as \imax~and the luminosity of satellites at the brightest and the faintest side of \gap are labelled as $M_{V, i}$ and $M_{V, i+1}$. Thus, for M101, it has \imax$=3$ with \Deltaimax$=5.4$, since its \gap is located between the third ($M_{V,3}$=-15) and the fourth ($M_{V,4}$=-9.6) satellites. 
   
\begin{figure}
   	\includegraphics[width=\linewidth]{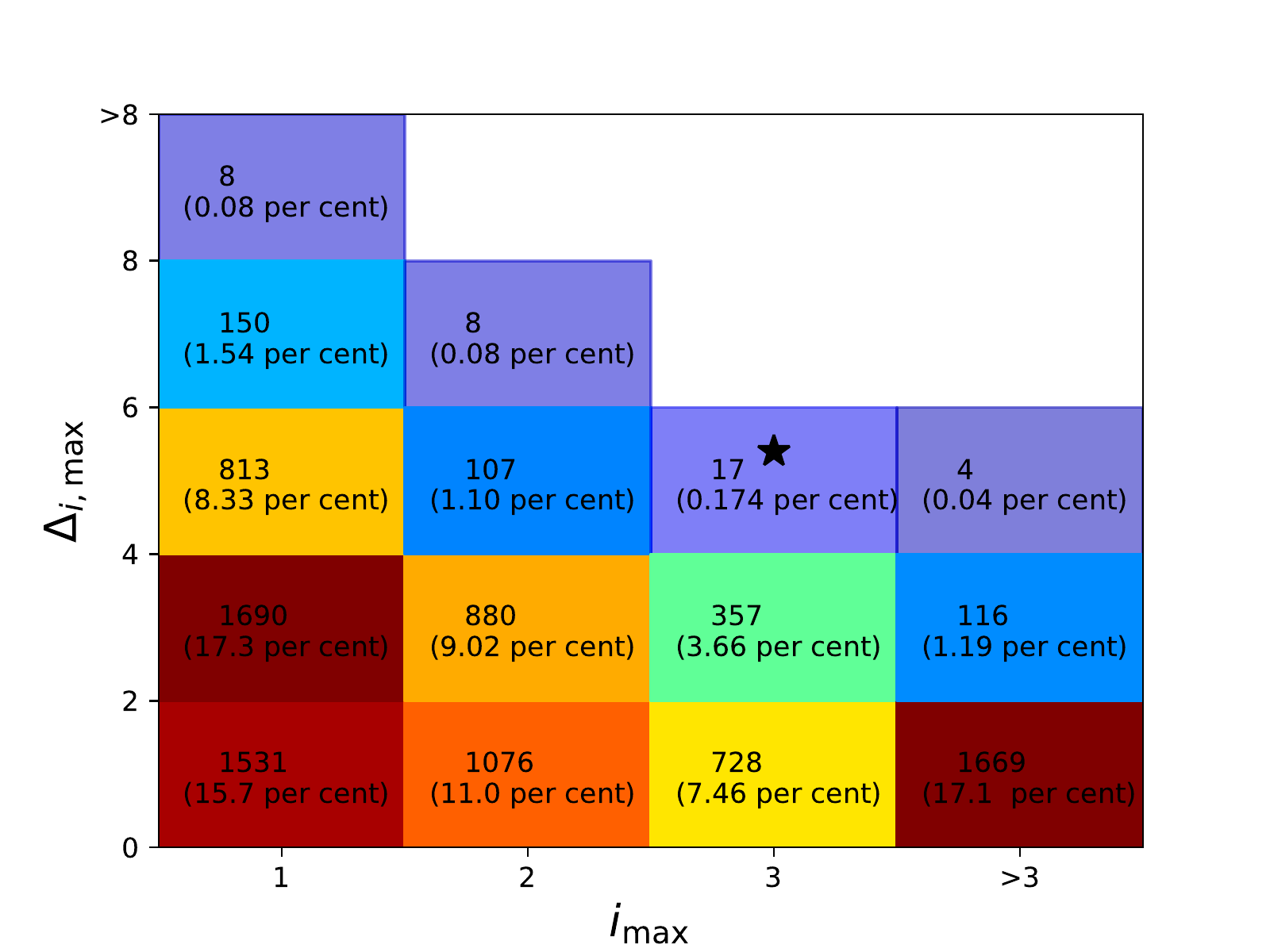}
   	\caption{The probability distribution of model galaxies in given \gap(\Deltaimax) and order of \gap(\imax) in SAM sample. The number count and number percentage of galaxy samples located in each region are labelled, where M101 is represented by the black star.}
   	\label{fig:table}
   \end{figure}

   We show the distribution of the \gap from our SAM data in Fig~\ref{fig:table}. There are about 11\% of SAM galaxies located at the region with \Deltaimax$>4$, and most of them have \imax$=1$. Only about 0.174\% of the model galaxies are located at a similar region as M101. If we take \Deltaimax$>$4 and \imax$=3$ as a standard choice of M101 analogous, it is quite hard to find M101-alike galaxies($\sim 3\sigma$) in our SAM sample.  
  
\begin{figure}
   	\includegraphics[width=\linewidth]{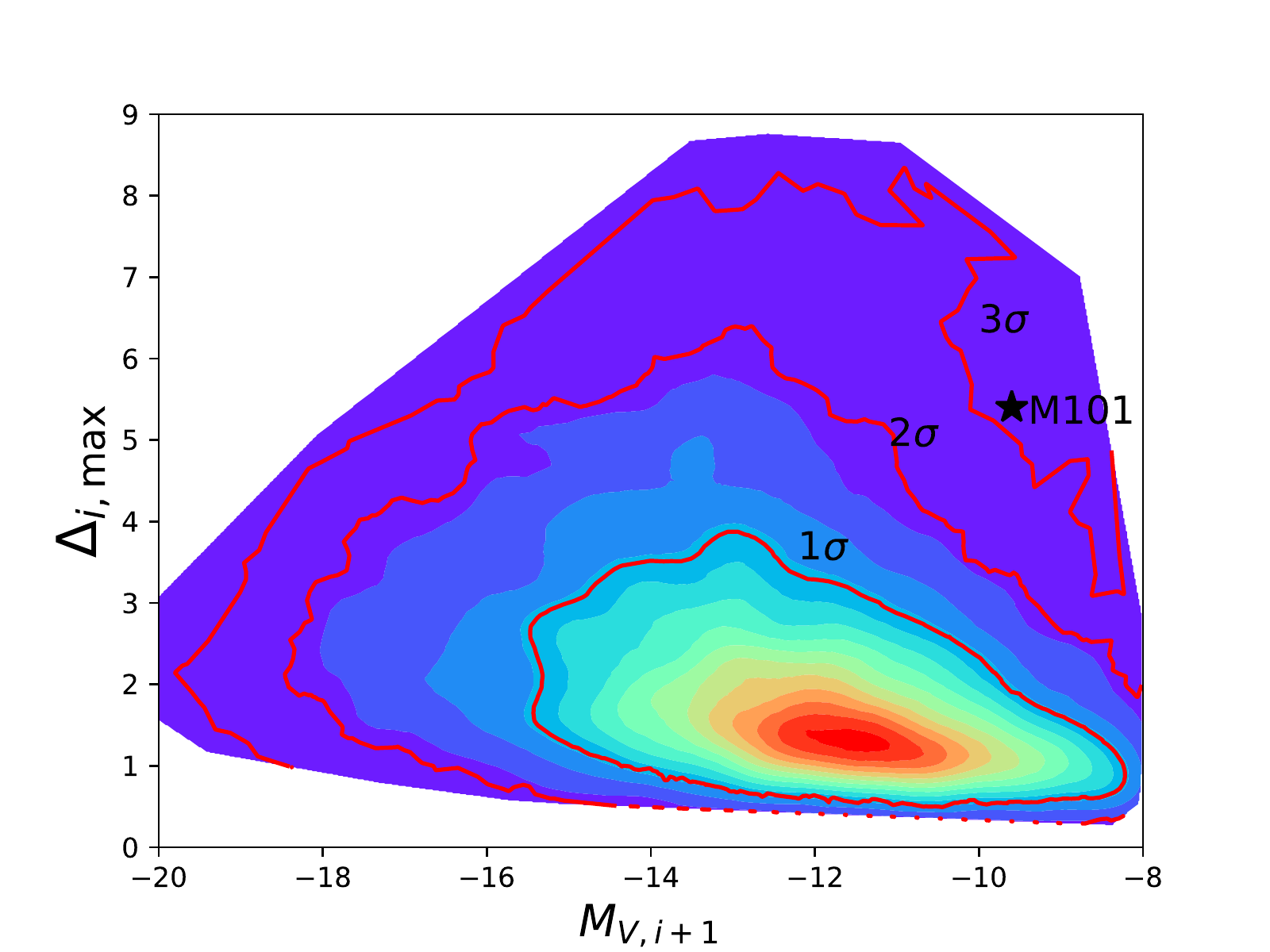}
   	\caption{The density contour distribution of galaxies in the \Deltaimax and $M_{V, i+1}$ space. The vertical axis represents the \gap in V band magnitude, the horizontal axis represents the magnitude of the satellite at the faintest side of the \gap. The contours, with 1$\sigma$, 2$\sigma$ and 3$\sigma$ region are shown as red lines. }
   	\label{fig:gap-pos}
   \end{figure}

 We note that in different host galaxies, the value of \gap may be the same, but the positions of \gap could be quite different, which could occur anywhere for satellites with different magnitudes. We use the magnitude at the faint end of the \gap, $M_{V, i+1}$, to represent the position of the \gap, and show the relationship between the value of \gap and the position of the \gap in Fig~\ref{fig:gap-pos}. The distribution between \Deltaimax ~ and $M_{V, i+1}$ shows a triangle like shape in the diagram. The M101 galaxy, marked as black star with \Deltaimax$= -5.4$ and $M_{V, i+1}=-9.6$, stays around the right edge of the distribution, with a larger \Deltaimax and fainter $M_{V, i+1}$ than most of the samples. There are still several model galaxies locating around M101 in the diagram. However, if we combine the restrictions in Fig~\ref{fig:table} and  Fig~\ref{fig:gap-pos} together to find M101-alike galaxies with similar \Deltaimax, $M_{V, i+1}$ and with $i_{max}=3$, there is only one model galaxy in our simulation data.
   
\begin{figure}
   	\includegraphics[width=\linewidth]{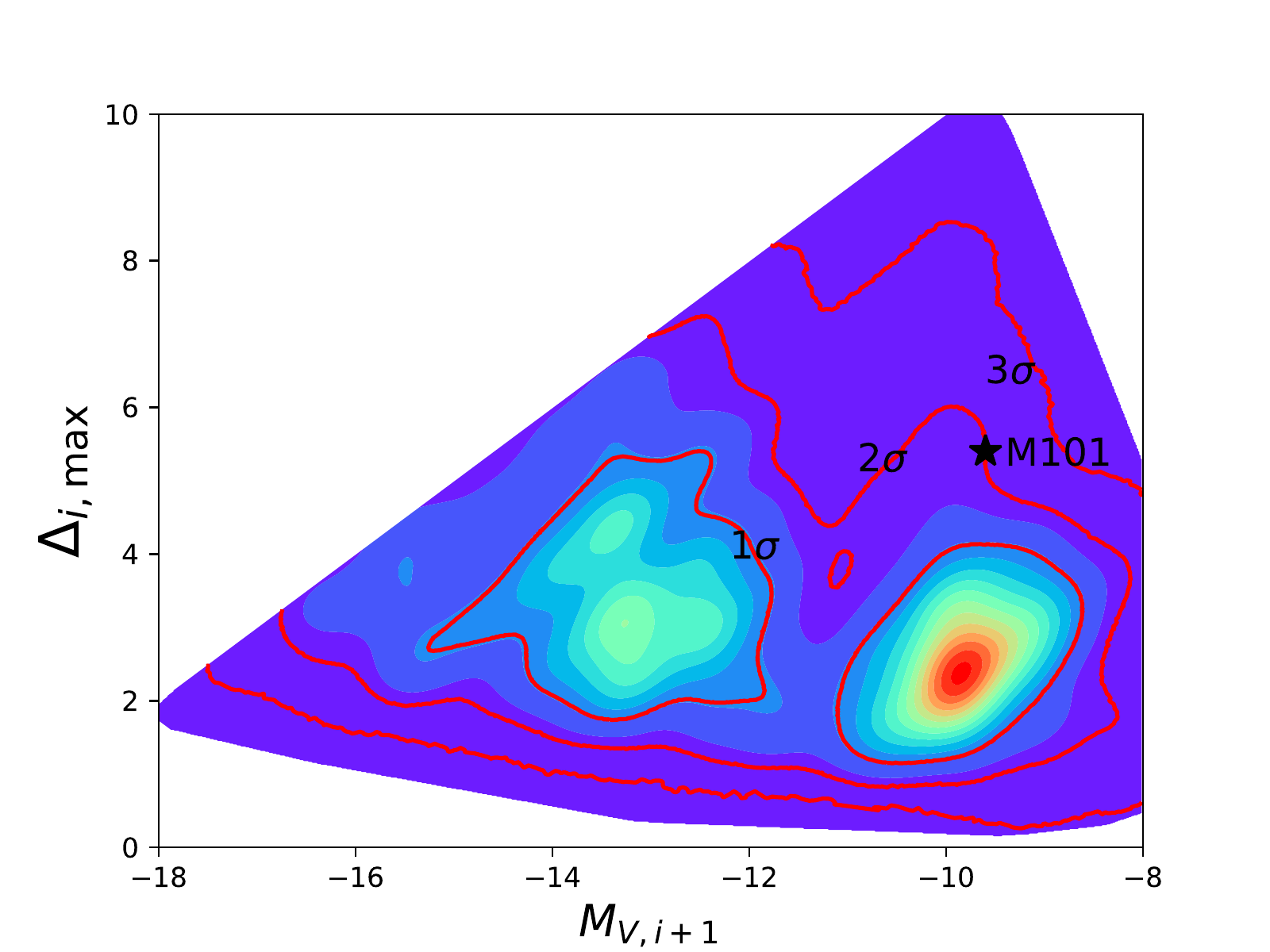}
    \caption{Similar as Fig. \ref{fig:gap-pos}, but it shows the distribution of the mock catalog from the Monte-Carlo realizations of the average luminosity function of satellites in a few observational MW-mass host galaxies (shown as the blue line in lower panel of Fig.~\ref{fig:sam-range}}
    \label{fig:gap-pos-mc}
   \end{figure}

 The above results show that M101 is an outlier in the  theoretical sample. To estimate if M101 is also an outlier in the observational data of MW-mass galaxies, we produce Monte Carlo realizations of the observational data in the MW-mass galaxies. To this end, we use the satellite luminosity functions from four observed MW-mass  galaxies (the color lines in Fig.~\ref{fig:sam-range}, not including the red line for M101) and obtain their average distribution (blue line in lower panel of Fig.~\ref{fig:sam-range}). We then generate one million Mote-Carlo realizations of the mean satellite luminosity functions in five host galaxies, and plot their distribution of the \gap in Fig.~\ref{fig:gap-pos-mc}.

 The distribution of \gap in the mock data is similar to the model prediction in Fig~\ref{fig:gap-pos}. There are two peaks at $M_V\sim-13$ and $M_V~-10$ in the mock data, which is due to the flat shape of the LF at $M_V<-12$ and $M_V<-10$. The black star represents M101, with \Deltaimax $=5.4$ and $M_{V, i}$=9.6, lying beyond 2$\sigma$ region of the diagram. Compared to the distribution in Fig.~\ref{fig:gap-pos}, the probability of finding a M101 galaxy from the observed MW-mass sample is slightly higher than the SAM prediction where it is around $3\sigma$ away from the peak. % this last sentence is not clear to me....
 Combining these results, we conclude that, although M101 is quite rare, we can find its analogous in both SAM model and the observed data. It implies that there might be a particular physical process leading to the formation of big gap in M101-alike galaxies in the CDM model. We will use the model data to investigate its origin in Sec.~\ref{sec:gap-origin}.
   
  \subsection{How is the \gap correlated with the galaxy halo mass?}
  As we have seen above, M101 is quite distinct from most other MW-mass galaxies in both simulation and observational data. Our previous work \cite{Zhang19} has shown that, for central galaxies with a given stellar mass, their halo mass are correlated with the mass of the most luminous satellite. Similarly, is there any dependence of halo mass on the \gap between satellites themselves? In this section, we investigate how the \gap is correlated with halo mass of the host galaxy.

  \begin{figure}
  	\includegraphics[width=\linewidth]{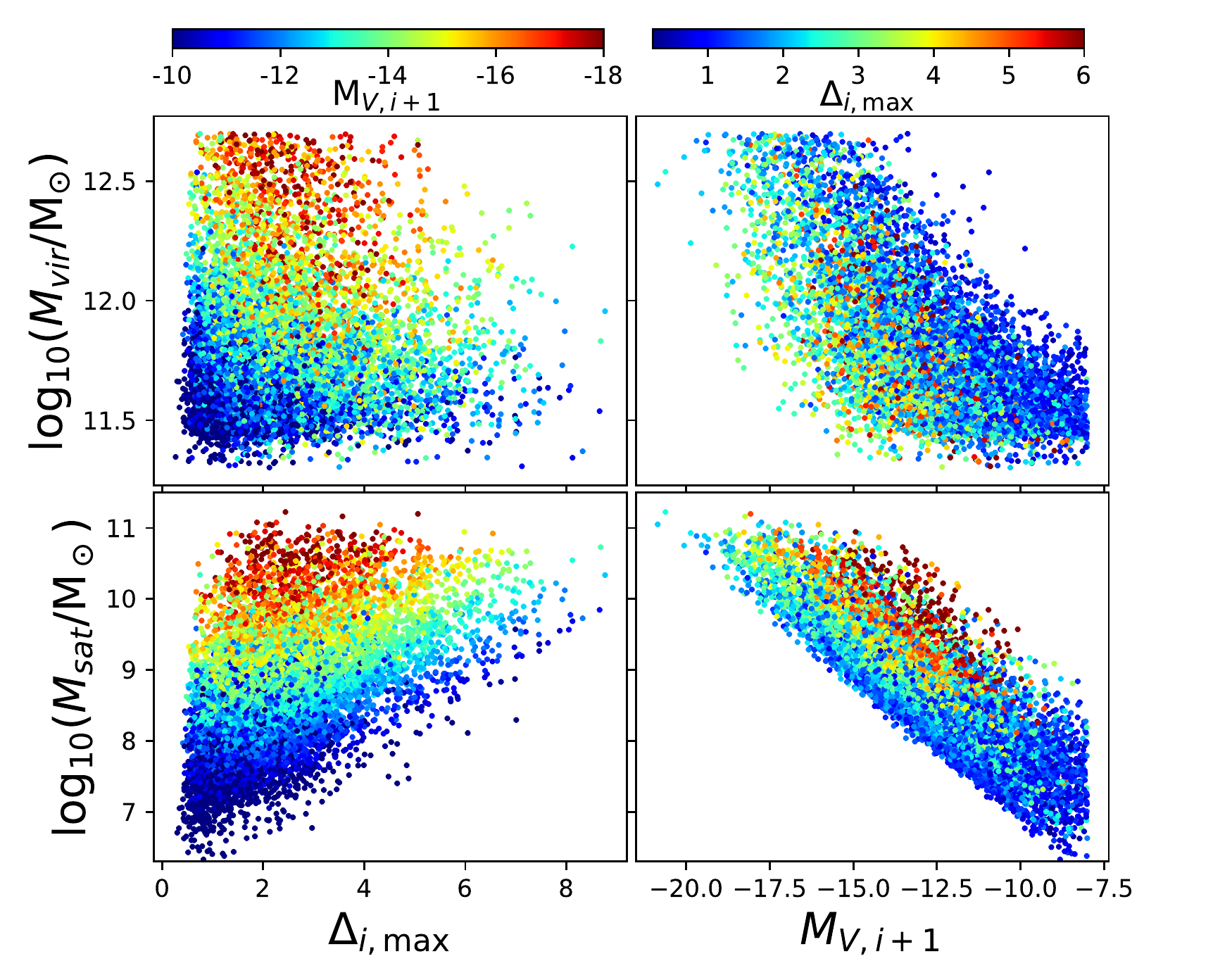}
  	\caption{The scatter distribution between the \gap and galaxy mass. The upper panels show the relation between gap and halo virial mass,  and the bottom panels are the distribution of gap and total satellite mass ($M_{sat}$). Every point represents a host galaxy}
  	\label{fig:gap-prop}
  \end{figure}
  
  We use the \gap value (\Deltaimax) and the \gap position ($M_{V, i+1}$) to fully specify the galaxy \gap. In Fig~\ref{fig:gap-prop}, we show the relation between the \gap and the halo mass or total stellar mass of the satellite galaxies. The upper panels show the relation between the \gap and halo virial mass. We can see from the left panel that, at a given \Deltaimax, galaxies have a wide distribution of host halo mass, showing weak dependence on \Deltaimax  alone. When galaxies are color-coded by $M_{V,i+1}$, a galaxy with larger $M_{V, i+1}$ (red color) has larger halo mass. This trend is more clear in the upper right panel where a strong dependence of the halo mass on $M_{V, i+1}$ can be seen. In the bottom panels, we show the distribution of the \gap and the total stellar mass of satellites. Again, similar to what is found in the upper panels, $M_{sat}$ has a strong correlation with $M_{V, i+1}$ (bottom right panel), but there is a very weak correlation between $M_{sat}$ and \Deltaimax, as shown in the bottom left panel. We also note another interesting feature in the left panels: for a host galaxy with larger \Deltaimax, the scatter of the halo mass or total stellar mass is smaller. As we will show in Sec.\ref{sec:gap-origin}, the gap is an indication of the formation time of the galaxy, such that host galaxies with big \gap have smaller scatter between their formation histories. %% this last sentence is not clear to me
  
  Therefore, compared with previous results on the gap between centrals and satellites as a strong indication of the host halo mass of a galaxy \citep{Dea13, Zhang19}, %%% I don't understand the meaning of this sentence....
  our results show that the dependence of the halo mass on the value of \gap between satellites is very weak, but the halo mass has a strong dependence on the luminosity of the satellites where the \gap between satellites occurs. %what do you mean with "happens"? Maybe "appears"?
  This indicates that the position of the \gap ($M_{V, i+1}$) can be a good indicator of the halo mass of a host galaxy. For example, we can use this result to predict the halo mass of M101. For the galaxy with $M_{V, i+1}$=-9.6, it is expected to have a halo mass of $\sim 4\times 10^{11} \rm M_\odot$ from the upper right panel. This estimation is consistent with the results of \citet{Tik15}, who showed that M101 is expected to have a halo mass between $2.04\times 10^{11} M_\odot$ and $7.5\times 10^{11} M_\odot$.

  \subsection{How did the M101-alike galaxies form?}
  \label{sec:gap-origin}
  
  % here I suggest significant change. Previous, we select M101 alike galaxies with $\Delta_{i,max}>4$ and $M_{v_i+1} <-11$. This is not right. We should strictly follow the definition of M101-alike galaxies. Based on Fig.3, I suggest to select M101 like galaxy as $\Delta{i,max}>4$ and $i=3$, and they should have more than 5 satellites (I guess you will have sample with around 17 galaxies. Because previous sample are dominated by galaxy with i=1 (also seen from Fig.3), so it means the gap is mainly due to infall of the most massive subhalo. But for M101 with i=3, the big gap formation may not be the same. So need to replace Fig.7 using new figure. You should plot the evolution of the gap for all the 17 galaxy, one line for one galaxy. Meanwhile, you can select M101-alike galaxies at z=1 and show the evolution of the gap for each galaxy
  \begin{figure}
  	\includegraphics[width=\linewidth]{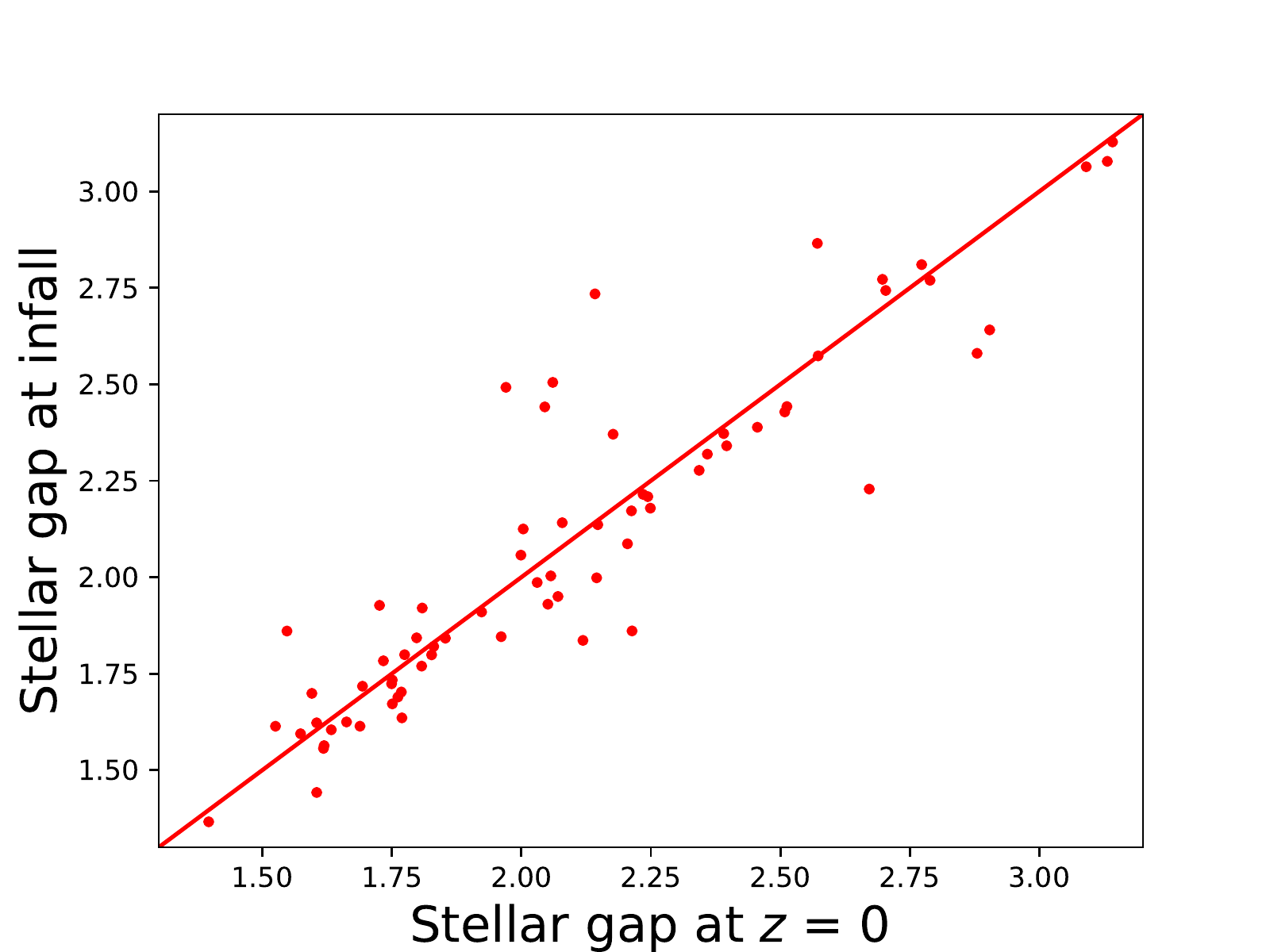}
  	%\caption{The gap in stellar mass of satellites at z=0 versus the gap in stellar mass at z_infall. }
    \caption{The \gap in stellar mass of satellites at z=0 versus the gap of those satellites at their infall times.}
  	\label{fig:gapzinfall}
  \end{figure}
  
  In this section, we investigate the formation of the M101-alike galaxies. As we have mentioned in Sec.2, by our definition, M101 has a gap of 5.4 which occurs between the third and fourth luminous satellite with a magnitude of $M_{V}=-9.6$. If we strictly follow this definition to find a M101-alike galaxy, we have only one model galaxy from the SAM data. To understand how M101-alike galaxies formed, here we select M101-alike galaxy as $i_{max}=3$ and with a gap larger than 3, in order to increase the size of the sample. For each M101-alike galaxy, we then trace its two satellite galaxies, which stay at the two sides of \gap, back to their infall times and record their stellar and halo mass at that time. Note that here we use the gap in stellar mass instead of the magnitude. For a constant mass-to-light ratio, the gap in stellar mass ($\Delta \lg M_*$) is 1/2.5 times the gap in magnitude. For example, the \gap in stellar mass of M101 is about 5.4/2.5=2.16. In Fig.~\ref{fig:gapzinfall} we show the distribution of the \gap in stellar mass at $z=0$ and that at the infall time of the satellites. There is a good correlation between the current \gap and that at infall time, indicating that the \gap we see today is linked to the \gap at infall time, and the evolution after infall is negligible. 

\begin{figure}
  	\includegraphics[width=\linewidth]{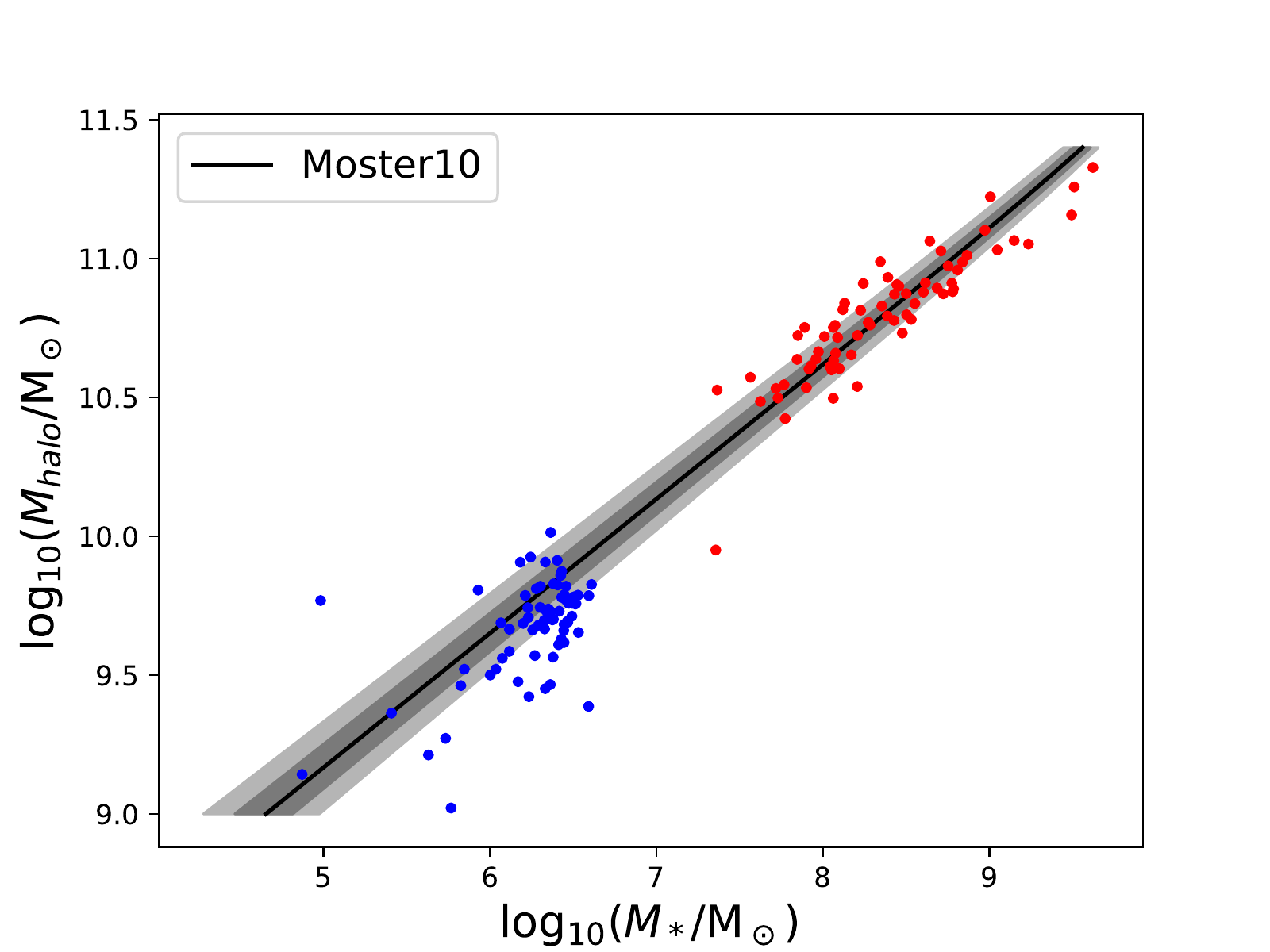}
  	\caption{The stellar mass-halo mass relation at the infall time for satellites at the two sides of the \gap. The red/blue points represent high/low mass satellites. }
  	\label{fig:halomasszinfall}
  \end{figure}

 To further study whether the big \gap is rooted in the halo mass of the satellites at infall time or the big \gap is originated from the stochastic star formation efficiency, in Fig.~\ref{fig:halomasszinfall} we show the relationship between the halo mass and stellar mass at infall time. The solid black line with grey shadow is from the halo abundance matching result at $z=0$ by \cite{Mos10}. The blue/red points are the satellites at the two sides of the \gap from SAM data, while the red points are for the massive satellites. We find that the \gap in stellar mass at infall is well correlated with the gap in halo mass (although with a slight larger scatter), and the predicted halo mass-stellar mass relation from the SAM is very similar to that from the halo abundance matching method at $z=0$. Also, the \gap in stellar mass is larger than that in halo mass, meaning that the star formation efficiency of satellite galaxies is halo mass dependent. The big \gap is not caused by the stochastic star formation in halos with similar mass, but dominated by the difference in halo mass at the time of infall. 

%better to include 2$\sigma$ scatter in the halo abundance matching relation in either Moster et al. 2009 or Leauthaud et al. 2011 in the right panel of Fig.?

At a first sight, our results seem to be inconsistent with that of \citet{Geh17}, who claimed that the halo abundance matching model cannot produce big \gap galaxies. Their Fig.15 shows that galaxies with \gap $\sim 5$ are beyond $2\sigma$ of the model predictions. In fact, one can read from their plot that these big-\gap galaxies can be actually covered within the $3\sigma$ distribution, similar to what we find from Fig.~\ref{fig:gap-pos} that M101-alike galaxies stay at around $3\sigma$ from the peak of the distribution. Our results indicate that M101-alike galaxies with \gap $\sim 5$ can be produced from the model, although with lower probability, and the \gap is rooted in the difference in the halo mass of the satellites at accretion.

\begin{figure}
  \includegraphics[width=\linewidth]{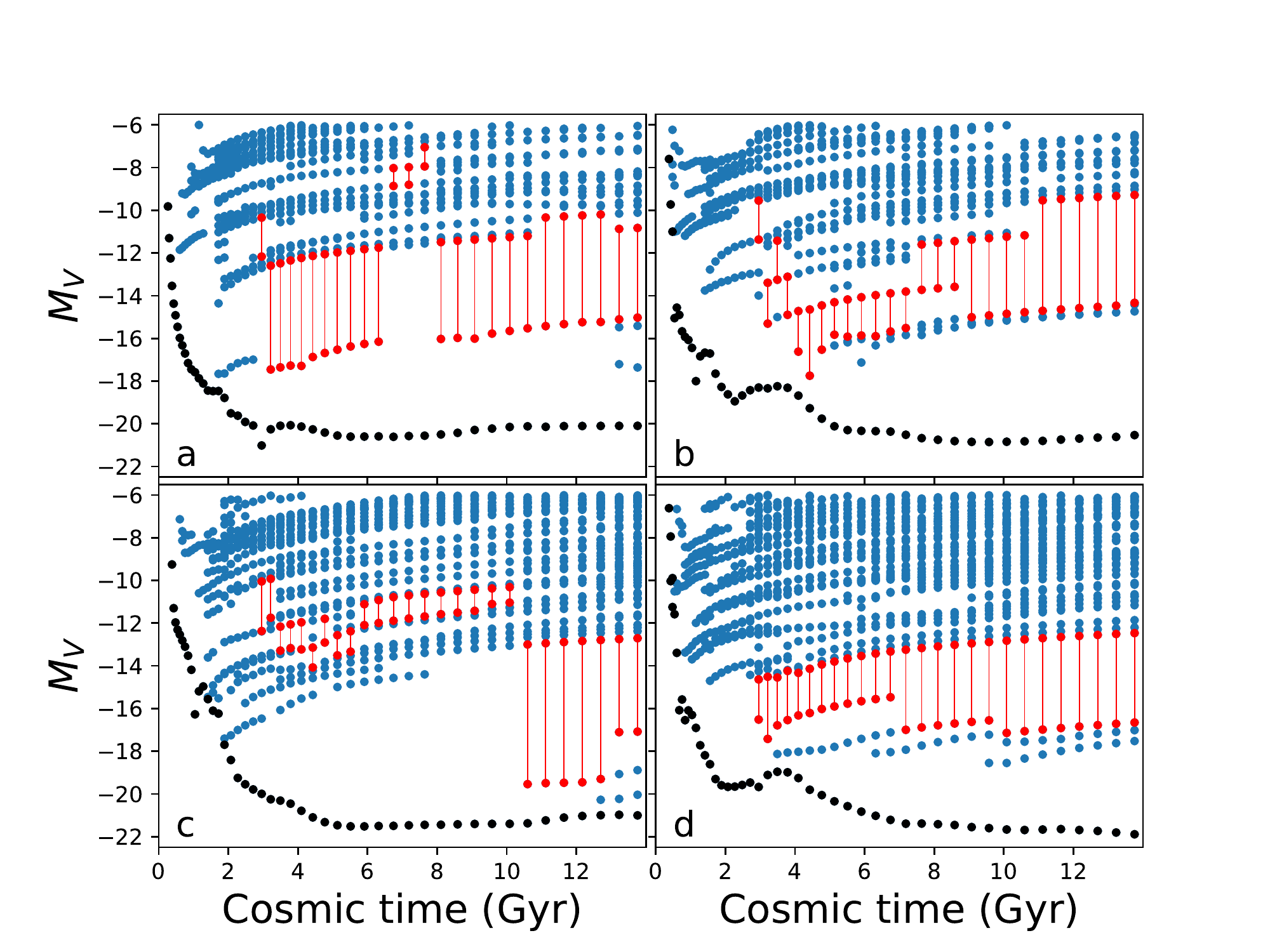}
  \caption{The magnitude evolution tracks for several M101-alike galaxies.  Central galaxies are represented by black dots and satellites are represented by blue dots. The satellites on the two sides of the \gap are represented by red dots and are connected by red lines.}
  \label{fig:sateevol}
\end{figure}
  
Our SAM has the advantage to trace the formation and evolution of the satellite galaxies. We find that the formation of the \gap in each M101-alike galaxy is very diverse and complicated, but in general it can be categorized into four cases, and we show some examples in Fig.~\ref{fig:sateevol}. 

% I would suggest to itemize the following

I) ``Stable'' case: Big \gap formed at high redshift and keep almost unchanged afterwards. As panel (a) shows, the satellites on the two sides of the \gap almost did not disappear (merge with central galaxy), and their magnitude gap evolved little. 

II) ``Death'' case: Big \gap formed at low redshift as the satellite at the faint side of the \gap disappeared at low redshift. As panel (b) shows, at high redshift, there are several bright satellites and the \gap was small, but during the recent ~4 Gyr, some bright satellites disappeared, by tidal disruption or merger with central galaxy, the faint side of the \gap was replaced by fainter satellite, and so the \gap became bigger.

III) ``Birth'' case: Big \gap formed at low redshift due to the new infall of bright satellites to form the bright side of the \gap. As panel (c) shows, at first the host galaxy lacked a brighter satellite, which means there was only a big gap between the central galaxy and the brightest satellite. During the recent ~4 Gyr, three bright satellites fall into the host galaxy, which lead to a birth of the \gap between satellites themselves.  

IV)``Combined'' case: Big \gap formation is not due to a single ``death'' or ``birth'' case, but from a combination of them.  As we show in the panel (d), the \gap evolution history is more complicated. The \gap growth is not only due to the disappearance of intermediate satellites, but also by the infall of bright satellites.

Another interesting feature we find is that, as shown in panel (c),  the recently accreted satellite happens to have lower luminosity than the previous satellite at the bright side of the \gap, and so it produces a smaller \gap. From the four cases shown in Fig.~\ref{fig:sateevol}, we could conclude that the luminosity of satellites did not change significantly, but the \gap is formed by "birth" or "death" of satellites at the two sides of the \gap, and the value of the \gap could increase or decrease. Overall, the formation of big \gap is mainly due to the recent accretion of satellites, not by their stochastic star formation efficiency.

\subsection{The evolution of the \gap}
We have seen in the previous section that the formation of the \gap in each host galaxy is diverse and complicated. To have a better understanding of how the \gap forms and evolves, we need to specify it in a statistical way. In particular, we want to know when the big-gap galaxies formed, how long the \gap can last, and whether the \gap is in a stable state. To increase our sample size, here we do not limit our sample to M101-alike galaxies, but we include those with a big-\gap (\Deltaimax$>3$), and we also do not take into account where the gap happens (no constraint on $i_{max}$). Here, we divide our model galaxies into three class. For sample A, we select MW-mass galaxies at $z=0$ with $M_V$ between -20 and -22. Sample B is a sub-sample of Sample A with \Deltaimax$>3$ at $z=0$. Sample C is selected from model galaxies at $z=1$ with \Deltaimax$>3$. For each sample, we then trace back to higher redshifts, and check the fraction of galaxies with \Deltaimax$>3$. For Sample C, we can also trace forth of its evolution to $z=0$ to study how long the gap can last. All the samples are shown in table~\ref{table1}.

 % In this section, we investigate the the formation process of galaxy groups with larger \gap like M101, by tracing back their formation history. Since the rarity of M101 analogous in our SAM data, we take a much wider standard to get more M101 analogous. Firstly, we take all 9756 MW-like galaxy groups in our work as control sample, represented as sample A. Depending on the value of the gap, we define the galaxies with a larger gap ($\Delta_{i, max}\ge$3) as sample B. To investigate the follow-up evolution of galaxies with large gap, we take the galaxies with $\Delta_{i, max}\ge$3 at z=1 as sample C. 
  
  \begin{table}
  	\centering
  	\begin{tabular}{|c|c|}
  		\hline
  		sample A & all MW-mass galaxy groups at $z=0$\\
  		\hline
  		sample B & sub-sample of A with \Deltaimax$\ge$3 \\
  		\hline
  		sample C & MW-mass galaxy with \Deltaimax$\ge$3 at $z=1$\\
  		\hline
  	\end{tabular}
  \caption{Selection of different samples used in this section.}
  \label{table1}
  \end{table}

\begin{figure}
  	\includegraphics[width=\linewidth]{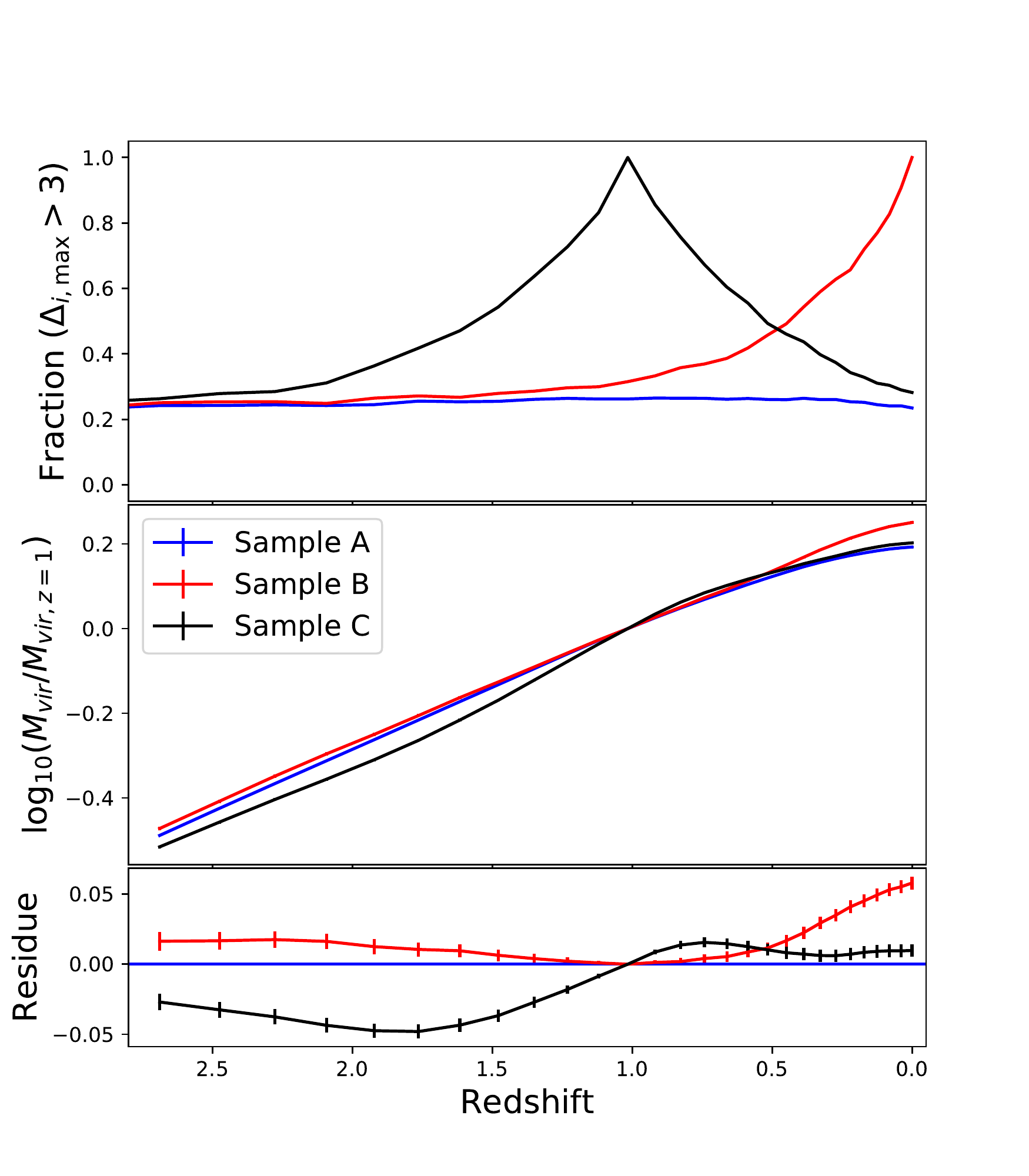}
  	\caption{The evolution of \Deltaimax and halo mass for selected galaxy samples. The upper panel shows the fraction of galaxies with \Deltaimax$\ge 3$ in each sample. The middle panel shows the growth of halo mass, normalized at $z=1$.  The residue of sample B and sample A, with respect to sample A is shown in the lower panel. }
  	\label{fig:gapevolution}
  \end{figure}

  In the upper panel of Fig~\ref{fig:gapevolution}, we show the fraction of galaxies with \Deltaimax$\ge$3 for each sample. We can see that for sample A, shown as blue line, the fraction of galaxies with \Deltaimax$\ge$3 has a relatively stable value of 20\% at different redshifts. For sample B, represented by the red line, the fraction of big \gap galaxies is quite similar to sample A before $z\sim1$. However, the fraction increases rapidly after $z\sim 0.5$. 
  It shows that the big-\gap (\Deltaimax$>3$) galaxies in Sample B formed in the recent $\sim$5 Gyrs. % I don't understand this last sentence...
   For those galaxies in sample C (selected with big \gap at $z=1$, the black line), the fraction starts to increase from $z=2$, and then decreases after $z=1$, and it becomes the same as Sample A at $z=0$. Results from Sample B and C show that galaxies with \gap with \Deltaimax$>3$ are not in a stable state across cosmological time scale, 
   but showing a mild evolution that the gap forms in the past 4$\sim$5 Gyrs and it can last as long as $\sim$7 Gyrs. % same....I don't understand this sentence....
  
  In Sec.\ref{sec:gap-origin}, we have found that the newly accreted satellite galaxies are the main origin for the big \gap, so it is reasonable to guess that the evolution of the \gap is linked to the halo formation history. To verify this, we compare the evolution of the halo mass from the three samples in the bottom panels of Fig~\ref{fig:gapevolution}. The middle panel shows the average halo mass evolution normalized at $z=1$, and the bottom panel the residue of sample B \& C with respect to sample A. As we can see, before $z=1$, Sample B shows similar halo mass growth rate as sample A, but it grows faster after $z=0.5$. Sample C increases its mass rapidly before $z=1$ and grows slower as Sample A after $z=1$. Compared with the trend in the top panel of Fig~\ref{fig:gapevolution}, we can see that the halo mass has consistent growth history as the \gap evolution. Thus, we can infer that the new accretion of satellites causes the growth of both the \gap and halo mass. As we discussed in Sec.\ref{sec:gap-origin}, both "birth" and "death" of satellites could form big \gap. The rapid growth of halo mass of big-\gap galaxies indicates that the evolution of the \gap is dominated by the "birth" case. 
  
 %The number increase of satellites on the bright end and \Deltaimax value should be induced by halo accretion. After z=1, sample 2 galaxies show similar grow curve with sample 0, on the same time, \Deltaimax value decrease. Which means the state of large \Deltaimax value is unstable, and these galaxies will evolve to a state with smaller \Deltaimax naturally, not depend on the halo accretion. So why these bright satellites disappear, and \Deltaimax value decrease without any special process? One might reason is that the lifetime of bright satellites is much shorter than fainter satellites, since bright satellites are locate in large subhalos which will lose their kinetic energy faster by dynamical friction. When these bright satellites exhaust their lifetime and merge into central galaxy, the big \gap between them vanish on the same time.
  
\begin{figure}
  	\includegraphics[width=\linewidth]{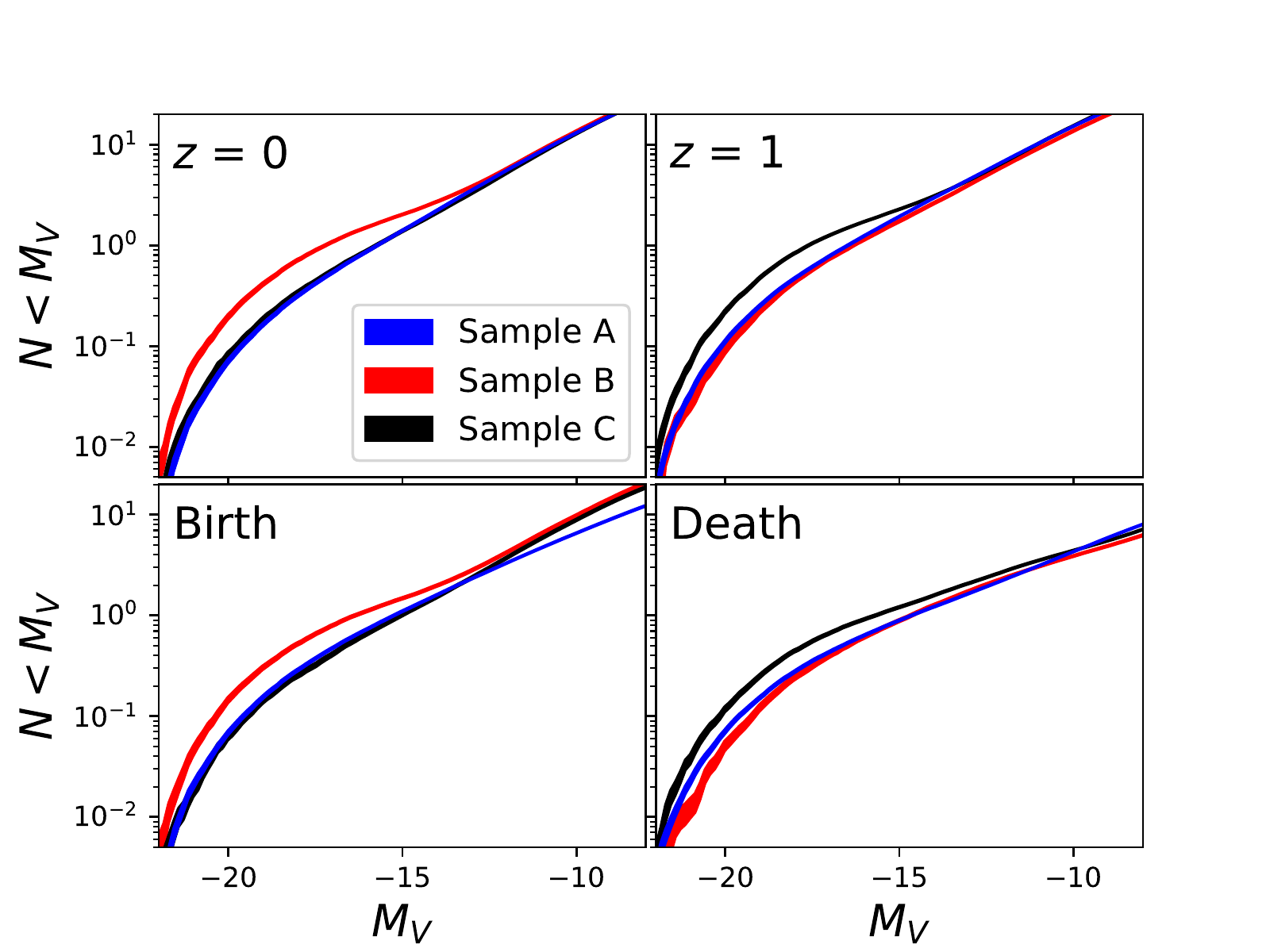}
  	\caption{The luminosity function of satellites for different samples. The upper panels show the average accumulated luminosity functions of satellites at $z=0$, and their progenitors at $z=1$. The lower panels show the average cumulative luminosity functions of satellites in ``birth'' and "death" mode between $z=1$ and $z=0$. See text for more details.}
  	\label{fig:redshift1-0}
  \end{figure}

  Now we study the evolution of satellite luminosity functions for the three samples at redshfit $1$ and $0$, to investigate which kinds of satellites contribute to the evolution of the \gap. In the upper panels of Fig~\ref{fig:redshift1-0}, we compare the average cumulative luminosity function of satellites for sample A, B and C, at $z=0$ (upper right panel) and their progenitors at $z=1$ (upper left panel). It is seen that for galaxies in Sample B (big \gap at $z=0$, the red line in the upper left panel), they have more bright satellites than Sample A at $z=0$, but their progenitors do not have more bright satellites at $z=1$ (upper right panel). The same is true for Sample C (with big \gap at $z=1$), which contains more bright satellites than Sample A at $z=1$ (black line in the upper right panel), but not different from Sample A at $z=0$. These results, again, confirm that galaxies with big \gap are mainly contributed by bright satellites. 
  
%  it and progenitors of sample C (the black line) at $z=1$ panel, the galaxies tend to contain more brighter satellites, while these is no significant difference compared to sample A at faint end, which means big \gap evolution is related to brighter satellites. The big \gap galaxies in sample C seems lose their brighter satellites after $z=1$. 
  
 Similar to our definition in Sec.~\ref{sec:gap-origin}, if a satellite can be found at $z=0$, while not at $z=1$, corresponding to new accretion case, we label it as ``birth'' mode. 
 Contrariwise, a satellite can be found at $z=1$, while not at $z=0$, no matter it is disrupted or merged into central galaxy, is labelled as  ``death'' mode.  % I don't understand this sentence....
 In the lower two panels of Fig~\ref{fig:redshift1-0}, we show the cumulative luminosity function of ``birth/death'' mode satellites from sample A, B and C.     
  
  For sample B galaxies, represented by red lines, there are more ``birth'' mode satellites at the brighter end than the other two samples (lower left panel). Similarly, for sample C galaxies, represented by the black lines, there are more ``death'' satellites on the brighter end. These results indicate that the newly accreted brighter satellites lead to the growth of big \gap and their disappearance reduce the gap. For all MW-mass  galaxies (sample A), the average value of \gap (\Deltaimax) is quite stable across different redshifts, indicating that creation and disappearance of big gap galaxies are in balance.

  \section{summary}
  
  The survey of nearby Milky Way-mass galaxies has found rich diversity of satellite galaxy population. % what does it mean?
  Among them, M101 is found to show a much bigger \gap in the luminosity of its satellite galaxies. Since the big \gap in M101 is quite rare in both  observation and theory prediction, it presents a challenge to the present model of galaxy formation. In this work, we used a semi-analytic model combined with a high resolution N-body simulation to investigate the properties of the \gap and its origin. We select central galaxies with similar V band luminosity as M101, and compared the properties of the galaxy system with big \gap to others. We also traced the formation history of our model galaxies to $z\sim 2.8$ to investigate the formation and evolution of the big \gap. Our results are summarized as followings: 
  
  \begin{itemize}
  \item A host galaxy like M101 with big \gap is very rare. Only about 0.1\% of the model galaxies and 0.2\% of the random sampling of observational data have similar \gap as M101, namely with $\Delta_{i, max}>4$ and $i_{max}>3$. Although the probability to find M101 is low, it can be predicted from the current galaxy formation model based on the cold dark matter simulation. 
  
  \item The position of \gap ($M_{V, i+1}$) has a good correlation with halo mass. From our SAM data, we predict that M101 has a halo mass of $\sim 4\times 10^{11}M_\odot$, which is smaller than most of the MW-alike galaxies with similar luminosity for central galaxies. This prediction could be tested in the future using observational data.
  
  \item The formation of a big \gap is mainly due to the halo mass of  satellites at accretion, not from the stochastic star formation efficiency in satellites.  
  
  \item The big \gap in a galaxy system is not in a fully stable state. Most galaxies with $\Delta_{i, max}>3$ are found to form large \Deltaimax value at $\sim$5 Gyrs ago, and the gap will last around $\sim$7 Gyrs. %Uhm...it forms 5 Gyr ago and lasts 7 Gyr? 
  The main origin of the big \gap is due to the recent accretion of bright satellites, consistent with the growth of the halo mass of the host galaxy.
  
  \end{itemize}
 
 The satellite population of nearby Milky Way-mass galaxies has long been the focus of the cold dark matter model. The state-of-the-art hydro-dynamical simulations \citep[e.g.,][]{Engler21} have successfully reproduced the diversity of the stellar mass function in the observational data. Our results have shown that the detail of the satellite mass distribution, such as the gap between the satellites, can also be reproduced in the frame of the cold dark matter model. A more interesting question is then to quantify the exact probability of finding any outlier from the real data.
  
  \section{Acknowledgements}
    We thank the referee for helpful suggestions and also thank Dr. Emanuele Contini for careful reading and correcting of the manuscript. The Millennium-II Simulation data bases used in this article and the web application providing online access to them were constructed as part of the activities of the German Astrophysical Virtual Observatory (GAVO). This work is supported by the  NSFC (No. 11861131006,11825303, 11703091,11333008), the 973 program (No. 2015CB857003) and cosmology simulation database (CSD) in the National Basic Science Data Center (NBSDC-DB-10, No. 2020000088). We acknowledge the science research grants from the China Manned Space project with NO. CMS-CSST-2021-A03, CMS-CSST-2021-A04.
    
  \section*{Data Availability}
    The data underlying this article will be shared on reasonable request to the corresponding author.

\end{document}